\documentclass{PoS}

\title{NLO and NNLO EWC for PV M{\o}ller Scattering}

\ShortTitle{NLO and NNLO EWC for PV M{\o}ller Scattering}

\author{Aleksandrs Aleksejevs  \\
        Memorial University, Corner Brook, Canada  \\
        E-mail: \email{aaleksejevs@grenfell.mun.ca   }}

\author{\speaker{Svetlana Barkanova}%
         \thanks{A footnote may follow.}\\
        Acadia University\\
        E-mail: \email{svetlana.barkanova@acadiau.ca}}

\author{Eduard Kuraev                                       \\
        Joint Institute for Nuclear Research, Dubna, Russia \\
        E-mail: \email{kuraev@theor.jinr.ru   }}

\author{Vladimir Zykunov  \\
        Belarussian State University of Transport, Gomel, Belarus\\
        E-mail: \email{vladimir.zykunov@cern.ch }}

\abstract{
High-precision electroweak experiments such as parity-violating  M{\o}ller scattering can provide indirect access to physics at multi-TeV scales and play an important complementary role to the LHC research program. However, before physics of interest can be extracted from experimental data, electroweak radiative corrections, which can significantly reduce the cross-section asymmetry, must be calculated with an unprecedented completeness and accuracy. Although the two-loop corrections are strongly suppressed relative to the one-loop corrections, they can no longer be dismissed for the upcoming precision experiments. We evaluate a full gauge-invariant set of one-loop and several types of two-loop radiative corrections for the parity-violating $e^-e^- \rightarrow e^-e^- (\gamma)(\gamma\gamma)$ scattering asymmetry by combining two distinct but mutually-reinforcing techniques: semiautomatic, precise, with FeynArts and FormCalc as base languages, and by hand, with some approximations. For 11 GeV relevant for the ultra-precise MOLLER experiment planned at JLab, the results obtained by two approaches are in excellent agreement, which gives us assurance that our calculations are error-free.
}

\FullConference{36th International Conference on High Energy Physics\\
                  4-11 July 2012\\
                  Melbourne, Australia}

\begin{document}

\section{Introduction}

Polarized M{\o}ller scattering has been a well-studied low-energy reaction for close 
to eight decades now \cite{M1932}, but recently has attracted especially active interest from both
experimental and theoretical communities as an excellent tool for measuring parity-violating (PV) asymmetries \cite{DM1979}.

The first observation of parity violation in M{\o}ller scattering was made by the E-158 experiment
at SLAC \cite{2}, which studied M{\o}ller scattering of 45- to 48-GeV polarized electrons on the
unpolarized electrons in a hydrogen target. Its result at low $Q^2$ = 0.026 $\mbox{GeV}^2$,
$A_{PV} = (1.31 \pm 0.14\ \mbox{(stat.)} \pm 0.10\ \mbox{(syst.)}) \times 10^{-7}$ \cite{E158} allowed one
of the most important parameters in the Standard Model (SM) -- the sine of the  Weinberg angle -- to be determined
with an accuracy of 0.5\% ($\sin^2 \theta_W$ = 0.2403  $\pm$ 0.0013 in the $\rm \overline{MS}$ scheme).
A very promising experiment measuring the $ep$ scattering asymmetry soon to be completed at JLab, 
Qweak \cite{QWeak},  aims to determine $\sin^2 \theta_W $ with relative precision of 0.3\%.
The next-generation experiment to study $e^-e^-$-scattering, MOLLER (Measurement Of a Lepton Lepton Electroweak
Reaction),  
planned at JLab following the 11 GeV upgrade, 
will offer a new level of sensitivity and measure the PV asymmetry in the scattering
of longitudinally polarized electrons off an unpolarized target to a precision of 0.73 ppb.
That would allow a determination of the weak mixing angle with an uncertainty of
$\pm 0.00026 \ \mbox{(stat.)} \pm 0.00013 \ \mbox{(syst.)}$ \cite{JLab12}, or about 0.1\%, an
improvement of a factor of five in fractional precision when compared with the E-158 measurement. At such precision, any inconsistency with the Standard Model will signal new physics, so M{\o}ller scattering experiments can provide indirect access to physics at multi-TeV scales and play an important complementary role to the LHC research program \cite{1}.

Obviously, before we can extract reliable information from the experimental data, it is necessary
to take into account higher-order effects, i.e. electroweak radiative corrections (EWC). The inclusion of EWC is an indispensable part of any modern experiment, but will be of the paramount importance for MOLLER. Of course, a significant theoretical effort has been dedicated to one-loop radiative corrections already. In \cite{ABIZ-prd}, we gave a short review of the literature and  calculated a full gauge-invariant set of the one-loop EWC
both numerically, with no simplifications, and by hand, in an approximate but compact form. 
The total correction was found to be close to $-70$\%, and we found no significant 
theoretical uncertainty coming from the largest possible source, the hadronic contributions to the vacuum polarization.
The dependence on other uncertain input parameters, like the mass of the Higgs boson, was below 0.1\%.

A much  larger theoretical uncertainty in the prediction for the asymmetry may come
from the two-loop corrections. According to \cite{Petr2003}, the higher-order corrections are suppressed by a factor of up to 5\%, depending on a type of corrections, relative to the one-loop result.  However, since the one-loop weak corrections for the M{\o}ller scattering are so large  and since the 11 GeV MOLLER experiment is striving for such an unprecedented precision, theoretical predictions for its scattering asymmetry must include  not only a full treatment of one-loop radiative corrections (NLO) but also leading two-loop correction (NNLO). 

One way to find some indication of a size of the higher-order contributions is to compare results obtained with different renormalization schemes.
Our calculations \cite{arx-2} in the on-shell and CDR (Constrained Differential Renormalization) schemes show a difference of about 11\%,
which is comparable with the difference of 10\% between
$\rm \overline{MS}$ \cite{Czar1996} and the on-shell scheme  \cite{Petr2003}.
It is also worth noting that although two-loop corrections to the cross section may 
seem to be small,  it is much harder to estimate their scale and behaviour for such a complicated observable as the
PV asymmetry to be measured by the MOLLER experiment.

We approach a formidable task of calcutating the NNLO EWC in stages. The two-loop EWC to the Born ($\sim M_0M_0^+$) cross section can be divided into two classes:
the $Q$-part induced by quadratic one-loop amplitudes ($ \sim M_1M_1^+$),
and the $T$-part corresponding to the interference of the Born and 
two-loop diagrams ($ \sim 2 \mbox{\rm Re} M_{2} M_{0}^+$).
The goal of this paper is to calculate the $Q$-part 
and a sub-set of the T-part including the gauge invariant set of boson self energies and vertices of two-loop 
amplitude $M_2$. Although it is too early to draw any final conclusions, we believe that the two-loop EWC may be larger that previously thought.

\section{General Notations and Matrix Elements}

The cross section of polarized M{\o}ller scattering
with the Born kinematics:
\begin{equation}
e^-(k_1)+e^-(p_1) \rightarrow e^-(k_2)+e^-(p_2),
\label{0}
\end{equation}
can be expressed as:
\begin{equation}
\sigma  = \frac{\pi^3}{2s} |M_0+M_1+M_2|^2 
\approx \frac{\pi^3}{2s} (M_0M_0^+ + 2 {\rm Re} M_1M_0^+ + M_1M_1^+ + 2 {\rm Re} M_2M_0^+),
\label{01}
\end{equation}
where
$\sigma \equiv {d\sigma}/{d \cos \theta}$ and 
$\theta$  is the scattering angle of the detected electron
with 4-momentum $k_2$ in the center-of-mass system of the initial electrons. The
 4-momenta of initial ($k_1$ and $p_1$) and final
($k_2$ and $p_2$) electrons generate a standard
set of Mandelstam variables:
\begin{equation}
s=(k_1+p_1)^2,\ t=(k_1-k_2)^2,\ u=(k_2-p_1)^2.
\label{stu}
\end{equation}
$M_0$, $M_1$ and $M_2$ are 
the Born (${\cal O}(\alpha)$),
one-loop (${\cal O}(\alpha^2)$) 
and two-loop (${\cal O}(\alpha^3)$) 
amplitudes (matrix elements), respectively.

The one-loop amplitude $M_1$ as a sum of boson self-energy (BSE),
vertex (Ver) and box diagrams.
We use the on-shell renormalization scheme from \cite{BSH86, Denner},
so there are no contributions from the electron self-energies.
The question of the dependence of EWC on
renormalization schemes and renormalization conditions (within the same scheme) was addressed 
in our earlier paper \cite{arx-2}.
Now we present the one-loop complex amplitude as the sum of IR and IR-finite parts $M_1 = M_1^\lambda + M_1^f$. The IR-finite part $M_1^f$ can be found in \cite{Q-part} and for the IR part we have:
\begin{equation}
M_1^\lambda = \frac{\alpha}{2\pi}  {\delta_1^{\lambda}} M_0,\
\delta_1^{\lambda} = 4 B \log\frac{\lambda}{\sqrt{s}},
\label{mmm1}
\end{equation}
where
$\lambda$ is the photon mass and 
the complex value $B$ can be presented in the following form (see, for example, \cite{KuFa}):
\begin{equation}
B= \log\frac{tu}{m^2s}-1 - i\pi.
\end{equation}
Analogously, the two-loop amplitude is the sum $M_2 = M_2^\lambda + M_2^f$, where
\begin{eqnarray}
M_2^\lambda = 
  \frac{\alpha}{2\pi}  {\delta_1^{\lambda}}  M_1^f
+ \frac{1}{8}  \bigl( \frac{\alpha}{\pi} 
 \bigr)^2 \bigl( 
{\delta_1^{\lambda}} \bigr)^2 M_0.
\label{mmm2} 
\end{eqnarray}
Note that the structure of first term in (\ref{mmm2}) is the same as in (\ref{mmm1}) in terms 
of the soft photon factorization.

Now we should make sure that the infrared divergences  are cancelled.
In a similar way used for the amplitudes, we present
the differential
cross sections as sums of $\lambda$-dependent
(IRD-terms) and $\lambda$-independent (infrared-finite) parts:
$ \sigma_1 = \sigma^{\lambda}_1 + \sigma^{f}_1,\ \
\sigma^V_{Q,T} =  \sigma^{\lambda}_{Q,T} + \sigma^{f}_{Q,T}$.
The one-loop cross section is already carefully evaluated with full control
of the uncertainties in \cite{ABIZ-prd}.
The simplest form for IRD-terms are:
\begin{equation}
\sigma^{\lambda}_{1} = \frac{\alpha}{\pi} \delta_1^{\lambda} \sigma_0,\
\sigma^{\lambda}_Q =  {\Bigl( \frac{\alpha}{2\pi} \Bigr)}^2
       \Bigl[   \delta_1^{\lambda} {\delta_1^{\lambda}}^*
      + 2 {\rm Re} \bigl( \delta_1^{f} {\delta_1^{\lambda}}^* \bigr) \Bigr] \sigma^0,\
\sigma^{\lambda}_T =
        {\Bigl( \frac{\alpha}{2\pi} \Bigr)}^2
    {\rm Re} \Bigl[ {\bigl( \delta_1^{\lambda} \bigr)}^2 + 2 \delta_1^{\lambda} \delta_1^{f} \Bigr] \sigma^0.
\end{equation}
The imaginary part of the total cross section cancels out in the sum $Q$- 
and $T$-parts due to following properties: 
$  \delta_1^{\lambda} {\delta_1^{\lambda}}^* + {\rm Re} {\bigl( \delta_1^{\lambda} \bigr)}^2
  =  2 {\bigl( {\rm Re} \ \delta_1^{\lambda} \bigr)}^2$,
and 
$ {\rm Re} \bigl( \delta_1^{f} {\delta_1^{\lambda}}^* \bigr) 
+  {\rm Re} \bigl( \delta_1^{f} {\delta_1^{\lambda}} \bigr) = 
 {\rm Re} (\delta_1^{f}) \ {\rm Re}  ({\delta_1^{\lambda}})$.
Thus, in
the following sections we can ignore the imaginary part, i. e. 
$ \delta_1^{\lambda} \rightarrow {\rm Re} \delta_1^{\lambda} $ and
$ B \rightarrow {\rm Re} B $.

\section{Bremsstrahlung and Cancellation of Infrared Divergences }

To evaluate the cross section induced by the emission of one soft photon with energy less then $\omega$,
we follow the methods of  \cite{HooftVeltman} (see also	\cite{KT1}).
Then, this cross section can be expressed as:
$ \sigma^{\gamma}=  \sigma^{\gamma}_1 + \sigma^{\gamma}_2$,
where $\sigma^{\gamma}_{1,2}$ have the similar factorized  structure
based on the factorization of the soft-photon bremsstrahlung:
$ \sigma^{\gamma}_{1,2}= \frac{\alpha}{\pi} \bigl[ -\delta_1^{\lambda} +R_1 \bigr] \sigma_{0,1}$,
where
\begin{equation}
R_1=-4B \log\frac{\sqrt{s}}{2\omega} - \Bigl( \log\frac{s}{m^2} - 1\Bigr)^2 +1-\frac{\pi^2}{3} +\log^2\frac{u}{t}.
\end{equation}
The first part of the soft-photon cross section,  $\sigma^{\gamma}_1$,  cancels the IRD at the one-loop order,
while the second part, $\sigma^{\gamma}_2$, cancels the IRD at the two-loop order,
with half of $\sigma^{\gamma}_2$ going to the cancellation of the IRD in the $Q$-part and the
other half going to treat IRD in the $T$-part.
At last, the cross section induced by the emission of two soft photons with a total energy less then $\omega$
is calculated in \cite{Q-part} as:
\begin{equation}
\sigma^{\gamma\gamma}=
\frac{1}{2}
{\Bigl( \frac{\alpha}{\pi} \Bigr)}^2
\bigl( \bigl( -\delta_1^{\lambda} + R_1 \bigr)^2 - R_2 \bigr)
\sigma_0,
\end{equation}
where $\frac{1}{2}$ is a statistical factor 
and $R_2 = \frac{8}{3}\pi^2 B^2$.

Combining all the terms, we get the infrared-finite result at both the
first and second orders:
\begin{equation}
\sigma_{\rm NLO} = \sigma_1 + \sigma^{\gamma}_1 = \frac{\alpha}{\pi} [R_1 + \delta_1^f] \sigma^0,
\label{1IR}
\end{equation}
\begin{eqnarray}
\sigma_{\rm NNLO} & =& \sigma_Q^V  + \sigma_T^V + \sigma_2^{\gamma}+ \sigma^{\gamma\gamma} 
=
{\Bigl( \frac{\alpha}{\pi} \Bigr)}^2
 [ R_1 \delta_1^f +\frac{1}{2} R_1^2 - \frac{1}{2}R_2 + \delta_Q^f + \delta_T^f  ]  \sigma^0 =
\nonumber
\\ & =& 
  \sigma_O^f + \sigma_B^f + \sigma_Q^f + \sigma_T^f,
\label{2IR}
\end{eqnarray}
where 
\begin{eqnarray}
  \sigma_O^f = \frac{\alpha}{\pi} R_1 \sigma_{\rm NLO},\ 
  \sigma_B^f = - \frac{1}{2} {\Bigl( \frac{\alpha}{\pi} \Bigr)}^2  (R_1^2 + R_2) \sigma^0.
\label{O-i-B}
\end{eqnarray}

\section{Numerical Results and Conclusions}

For the numerical calculations at the central kinematic point of MOLLER
($E_{lab}$=11 GeV, $\theta =\pi/2$)
we use
$\alpha$,\ $m_W$,  $m_Z$ and lepton masses as input parameters in accordance with \cite{PDG08}.
The effective quark masses which we use for the vector boson self-energy loop contributions are extracted from shifts in the 
fine structure constant due to hadronic
vacuum polarization $\Delta \alpha_{had}^{(5)}(m_Z^2)$=0.02757 \cite{jeger}.
For the mass of the Higgs boson, we take $m_H=125\ \mbox{GeV}$ 
and for the maximum soft photon energy we use $\omega = 0.05\sqrt{s}$, 
according to   \cite{ABIZ-prd} and \cite{5-DePo}.

Let us define the relative corrections to the Born cross section due to a 
specific type of contributions (labeled by $C$) as
$$\delta^{C} = (\sigma^{C}-\sigma^0)/\sigma^0,\ \ C=\mbox{NLO}, O, B, Q, T, \mbox{NNLO}.$$
In the text below the term "$T$-part" corresponds to the contributions of a gauge invariant set of the BSE
and vertices only. The parity-violating asymmetry is defined in a traditional way:
\begin{equation}
A_{LR} =
 \frac{\sigma_{LL}+\sigma_{LR}-\sigma_{RL}-\sigma_{RR}}
      {\sigma_{LL}+\sigma_{LR}+\sigma_{RL}+\sigma_{RR}}
 =
 \frac{\sigma_{LL}-\sigma_{RR}}
      {\sigma_{LL}+2\sigma_{LR}+\sigma_{RR}},
\label{A}
\end{equation}
and the relative corrections to the Born asymmetry  due to  $C$-contribution
are defined as
$$\delta^{C}_A = (A_{LR}^{C}-A_{LR}^0)/A_{LR}^0.$$

Our numerical estimations are presented in the table below, with dots in 4-th and 7-th columns denoting
the result written in previous line:

\bigskip

{\small
\hspace{-8mm}
\begin{tabular}{|c|c|c||c|c|c||c|c|c|}
\hline
  \multicolumn{1}{|c|} {~~~~$C$~~~~} 
& \multicolumn{1}{ c|} {$\delta^C$  }
& \multicolumn{1}{ c||} {$\delta_A^C$}
& \multicolumn{1}{ c|} {~~~~$C$~~~~}          
& \multicolumn{1}{ c|} {$\delta^C$  }
& \multicolumn{1}{ c||} {$\delta_A^C$}
& \multicolumn{1}{ c|} {~~~~~$C$~~~~~}          
& \multicolumn{1}{ c|} {$\delta^C$  } 
& \multicolumn{1}{ c|} {$\delta_A^C$} \\
\hline 
NLO & $  -0.1144 $ & $  -0.6932 $ & NLO     & $  -0.1144 $ & $ -0.6932 $ & NLO           & $ -0.1144 $ & $  -0.6932 $  \\
$O$ & $ ~ 0.0457 $ & $ ~ 0.2347 $ & ...+$O$ & $  -0.0687 $ & $ -0.3956 $ & ...+$(O+B)/2$ & $ -0.1339 $ & $  -0.5671 $  \\
$B$ & $  -0.0848 $ & $ ~ 0      $ & ...+$B$ & $  -0.1535 $ & $ -0.4353 $ & ...+$Q$       & $ -0.1150 $ & $  -0.6392 $  \\
$Q$ & $ ~ 0.0189 $ & $  -0.0731 $ & ...+$Q$ & $  -0.1345 $ & $ -0.5118 $ & ...+$(O+B)/2$ & $ -0.1345 $ & $  -0.5118 $  \\
$T$ & $ ~ 0.0119 $ & $  -0.1063 $ & ...+$T$ & $  -0.1226 $ & $ -0.6274 $ & ...+$T$       & $ -0.1226 $ & $  -0.6274 $  \\
\hline
\end{tabular}
\vspace{5mm}
}

As one can see from our numerical data, at the MOLLER kinematic conditions, the part of
the NNLO EWC we considered in this work can increase the asymmetry by up to $\sim$ 7\%. 
The $Q$- and $T$-parts do not cancel each other but, on the contrary, are adding up to increase the physical PV effect.
Clearly, the large size of the investigated parts demands a detailed and consistent consideration of
the rest of the $T$-part, which will be the next task of our group. Since the problem of EWC for the
M{\o}ller scattering asymmetry is rather involved, a tuned step-by-step comparison between different
calculation approaches is essential. To make sure that our calculations are error-free, we control our results by comparing the data obtained from the equations
derived by hand with the numerical data obtained with a semi-automatic approach based on FeynArts,
FormCalc, LoopTools and Form. These base languages have already been successfully employed
in similar projects (\cite{ABIZ-prd}, \cite{arx-2}), 
so we are highly confident in their reliability. In the future, we plan
to address the remaining two-loop electroweak corrections up to a level required by the plannedl precision of the MOLLER experiment and the possible future experiments at ILC.

\section{ACKNOWLEDGMENTS}

We are grateful to Yu. Bystritskiy and T. Hahn
for stimulating discussions.
A. A. and S. B. thank the Theory Center at Jefferson Lab, and V. Z. thanks
Acadia University for hospitality. This work was supported by the
Natural Sciences and Engineering Research Council of Canada
and Belarus scientific program "Convergence".

\end{document}